# MRI denoising using Deep Learning

# and Non-local averaging

José V. Manjón[1] and Pierrick Coupe[2,3]

jmanjon@fis.upv.es, pierrick.coupe@labri.fr

[1]*Instituto de Aplicaciones de las Tecnologías de la Información y de las Comunicaciones Avanzadas (ITACA), Universitat Politècnica de València, Camino de Vera s/n, 46022 Valencia, Spain.*

[2] *Univ. Bordeaux, LaBRI, UMR 5800, PICTURA, F-33400 Talence, France.*

[3] *CNRS, LaBRI, UMR 5800, PICTURA, F-33400 Talence, France.*

* Corresponding author: José V. Manjón. Instituto de Aplicaciones de las Tecnologías de la Información y de las Comunicaciones Avanzadas (ITACA), Universidad Politécnica de Valencia, Camino de Vera s/n, 46022 Valencia, Spain

Tel.: (+34) 96 387 70 00 Ext. 75275, Fax: (+34) 96 387 90 09

E-mail: jmanjon@fis.upv.es (José V. Manjón)

**Abbreviations:**

CNN: Convolutional Neural Network

NLM: Non Local Means

PSNR: Peak signal to Noise Ratio

RMSE: Root mean squared error

SSIM: Structural similarity index

**Abstract**

*Purpose:*

This paper proposes a novel method for automatic MRI denoising that exploits last advances in deep learning feature regression and self-similarity properties of the MR images.

*Methods:*

The proposed method is a two-stage approach. In the first stage an overcomplete patch-based convolutional neural network blindly removes the noise without specific estimation of the local noise variance to produce a preliminary estimation of the noise-free image. The second stage uses this preliminary denoised image as a guide image within a rotationally invariant non-local means filter to robustly denoise the original noisy image.

*Results:*

The proposed approach has been compared with related state-of-the-art methods and showed competitive results in all the studied cases while being much faster than comparable filters.

*Conclusion:*

We present denoising method that can be blindly applied to any type of MR image since it can automatically deal with both stationary and spatially varying noise patterns.

# 1. Introduction

Magnetic Resonance (MR) images play a major role in current medical and research settings. Unfortunately, these images are inherently noisy due to its acquisition process and thus filtering methods are usually required to improve their quality.

There is a vast amount of bibliography related to this topic which highlights its relevance to the scientific community [1]. Before the rise of the so-called *deep learning*, most denoising methods were classified on two wide categories; those using the intrinsic pattern redundancy of the image patterns and those exploiting their sparseness properties.

In the first category, the proposed methods reduce the image noise by exploiting the self-similarity of the image patterns by averaging similar image patterns (typically image patches). Probably, the best example of this kind of methods is the well-known non-local means (NLM) filter [2]. In MRI, early works using the NLM method were proposed by Coupe et al [3] and Manjón et al. [4]. Numerous extensions and incremental improvements of these methods have been proposed in the past years [5-12].

In the second category, sparseness-based methods aim to denoise the images by assuming that the noise-free patterns can be sparsely represented in a lower dimensionality space while random noise patterns cannot be compressed. Such signal-noise separation can be performed in the image domain or into a transformed space (using for example FFT or DCT). For example, in [13] noise reduction was achieved removing DCT coefficients related to noise at patch level using either soft or hard thresholding techniques in natural images. For MRI, Manjón et al. [7] proposed a similar approach using 3D patches with pseudo-oracle strategy to robustly estimate the DCT coefficients to remove. In these approaches, overcomplete patch denoising was performed which significantly reduced block and Gibbs artifacts.

At image domain, methods learning image specific bases [14-16] where able to outperform the use of standard bases (such as DCTs) due to a much better pattern compression ratio. These techniques learn a set of bases from the images to be denoised to create a dictionary to sparsely represent image patches as a linear combination of dictionary atoms [17]. Although this dictionary learning can be performed offline (for same image type), usually online learning provides better results thanks to a better image content adaption [18].

An efficient manner to extract a sparse representation of a set of patches is using Principal Component Analysis (PCA) [19-21]. Using PCA, the original signal can be projected into an orthogonal space where most of the variance of the signal is accumulated in the first components while the noise being not sparse is uniformly spread over all the components. Noise reduction is normally achieved removing noise-related components and inverting the PCA decomposition [22-24]. In MRI denoising, PCA has been also used in numerous methods [6,25,26].

More recently, hybrid method using principles of both categories have been proposed. For example, in [18] used non-local patch matching to obtain a more compact group of patches prior the application of a PCA decomposition which allowed to obtain a very sparse representation. In a different manner, Manjón et al. [7,8] proposed two methods that *prefilter* the noisy image using sparse theory (DCT and PCA respectively) before using this prefiltered image as a guide image to finally filter the original noisy image using non-local similarities through a rotational invariance version of the non-local means filter.

Recently, deep learning methods have been proposed to denoise natural images using different architectures [27,28]. Most of these methods use supervised learning by training different architectures with pairs of noisy and noise-free input and outputs respectively. Such learning-based methods try to infer the clean image from the noisy input. One of the main benefits of these techniques is that, after training, the denoising can be applied extremely fast (on GPUs).

One of the first deep learning methods for medical image denoising was proposed by Gondara [29] using a convolutional denoising autoencoder though a bottleneck strategy to denoise 2D images. In [30] Benou et al. proposed a spatio-temporal denoising method using restricted Boltzman machines. More recently, Jiang et al. [31] proposed a specific Rician noise filter using a slice-wise convolutional neural network.

In this paper, we present a hybrid denoising approach based on the application of a 3D Convolutional Neural Network (CNN) using an overcomplete patch-based sliding window scheme. The obtained *prefiltered* image is used as a guide image to accurately estimate the voxel non-local similarities that are used by a rotationally invariant NLM (RI-NLM) to perform the final denoising as previously done in Manjón et al. [7,8]. This is an extension of a preliminary work previously presented at the MICCAI conference [32] were we have slightly modified the network architecture and extended the analysis of proposed method including also the spatially varying noise case.

## 2. Methods

Let define a noisy image *Y* as the original noise free signal *A* plus some noise *N*:

$$Y = A + N \tag{1}$$

The aim of any denoising method is to estimate A given *Y*. In MRI, the noise present in the images can follow different distributions depending on the acquisition process. The simplest approach is to assume that the noise follows a Gaussian distribution with zero mean and a given standard deviation $\sigma$. However, this is only true for single coil acquisitions and for local signal to noise (SNR) ratios higher than 3 [33]. The most common model for single coil acquired images is the Rician model [34] which is a non-zero mean signal dependent noise. In the last years, multi-coil acquisitions such as SENSE or GRAPPA have become quite common in clinical and research settings. Although the use of these sequences has reduced significantly the acquisition time this comes at the expense of more complex, spatially varying noise patterns [35].

In this paper we present a novel method based on the application of a patch-based CNN trained with noisy and their corresponding noise free versions.

### 2.1. Experimental data

***Training/validation dataset***

To train a supervised neural network ground truth data are needed. However, zero noise images do not exist and therefore we used real denoised images as surrogates of noise-free images.

We used images from the IXI dataset (http://www.brain-development.org). This dataset contains images of nearly 600 healthy subjects from several hospitals in London (UK). Both 1.5 T and 3 T images were included in our training dataset. 3T images were acquired on a Philips Intera 3T scanner (TR = 9.6 ms, TE = 4.6 ms, flip angle = 8°, slice thickness = 1.2 mm, volume size = 256 × 256 × 150, voxel dimensions = 0.94 × 0.94 × 1.2 mm$^3$). 1.5 T images were acquired on a Philips Gyroscan 1.5T scanner (TR = 9.8 ms, TE = 4.6 ms, flip angle = 8°, slice thickness = 1.2 mm, volume size = 256 × 256 × 150, voxel dimensions = 0.94 × 0.94 × 1.2 mm$^3$). All images were previously filtered using a PRI-NLPCA filter which represents the current state-of-the-art in MRI denoising [8].

Specifically, we randomly selected 5 denoised T1 MRIs from this dataset at every epoch using a generator to feed the network. Denoised images had virtually almost zero noise and the anatomy was minimally affected by the application of the filter. To train the network several levels of stationary Gaussian noise (range from 1% to 9%) were added to generate the training data. To create the validation dataset, we used the same approach using a fix set of 5 T1 MRIs out of the training dataset.

***Test dataset***

To be able to quantitatively compare the proposed method with previous methods, we used the well-known Brainweb T1 3D MRI phantom [36] as test dataset. This synthetic dataset has a size of 181x217x181 voxels (1 mm$^3$ voxel resolution) and was corrupted with different levels of stationary and spatially varying Gaussian and Rician noise (1% to 9% of maximum intensity). Rician noise was generated by adding Gaussian noise to real and imaginary parts and then computing the magnitude image.

A qualitative analysis was performed on 2 real MR images in which significant noise was induced by acquisition at very high resolution (one T1 and one T2 acquired on a Canon 3T ZGO scanner with high gradient performances) to evaluate the effectiveness of the proposed method on real conditions. The T1 was a SPEEDER (factor=1.6) MPRAGE (Fast Field Echo 3D) : TR: 7,2ms  -  TE: 3,2ms  -  Flip Angle: 9° -  TI: 900ms  - FOV: 22,4 x 22,4 cm  -  Matrix: 448 x 448  -  Resolution: 0,5 x 0,5 x 0,6 mm$^3$  Slices : 256. The T2 was a SPEEDER (factor=2) 3D T2 MPV : TR: 3000ms   - TE: 341,6ms (Echo Space 6,5)  -  FA: 90°/180°  -  FOV: 22,4 x 22,4 cm  -  Matrix: 448 x 448  - Resolution: 0,5 x 0,5 x 0,6 mm$^3$  Slices : 256.

## 2.2. Preprocessing

A common preprocessing step when using CNNs consists of intensity normalizing the input data. This is normally done centering the images by subtracting the mean and dividing by the standard deviation. Since our proposed method uses 3D patches as input of the network, this operation could be done to each patch independently. However, since we use a sliding window approach to denoise the images, we used a different approach to minimize block artifacts that could arise after mean and standard deviation restoration. First, we estimated a low-pass filtered image with a box-car kernel of 6x6x6 voxels (local mean map). Second, we estimated local standard deviation map using the same patch size (local standard deviation map). Afterwards, these two images were used to normalize the input and output volumes by subtracting the local mean map and dividing by the local standard deviation map. We found that this approach introduces significantly fewer blocking artifacts than the standard approach. Note that after this preprocessing most of the anatomic information is removed (remaining mostly edge information).

## 2.3. Proposed method

The proposed approach is based on a 3D patch-based convolutional neural network (PBCNN). The input and output of the proposed CNN are 3D patches of size *12x12x12* voxels. Such patches are extracted from the pre-processed images (as previously described) in an overcomplete manner with an overlapping of half patch (6 voxels) in all three dimensions. The topology of the proposed network is the following (see figure 2).

First, one input block of size 12x12x12 composed of one 3D convolution and a RELU layer. Then, seven repeated blocks composed of an Instance-Normalization, a 3D convolution and a RELU layers (we used 7 blocks to have receptive filed covering the whole input patch). Finally, a last block composed of an Instance-Normalization and a 3D convolution layer that produces a 12x12x12 output patch (see figure 1). We used InstanceNormalization layer instead of the usual BatchNormalization layer to retain the original variance of each patch of the mini batch which produces slightly better results. All convolution layers have 64 filters of 3x3x3 voxels. The whole network has a total of 779,009 trainable parameters. To train the network, we used ADAM optimizer, 300 epochs and a batch size of 128 patches. We used an early stop criterion using the validation data.

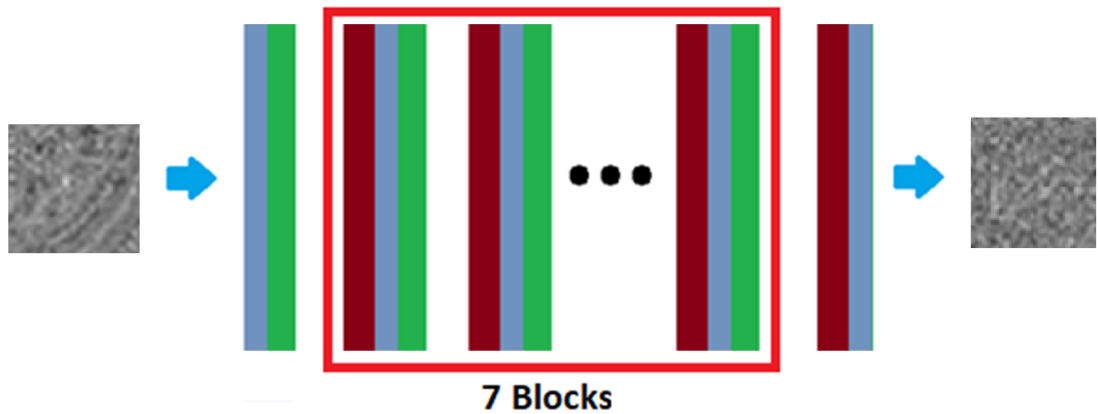

*Figure 1*. *2D example of the proposed patch-based CNN model. Block design: Red (InstanceNormalization layer), Blue (3D convolution layer of 64 filters of 3x3x3 voxels) and Green (RELU layer).*

We used a residual approach (i.e., the network learns how to remove the signal to produce the noise map) as in [31] instead of removing the noise from the input patch to produce a clean denoised patch. We found this option more effective (faster training and better results). Basically, instead of learning the noise-free patch, we learn the noise present in the patch. This is done by the network by removing correlated information in the input layer. Differently from [31] where the network removes the original image from the input, our network solves a simpler problem thanks to its specific preprocessing. Besides, also differently from [31] were different networks were trained to filter different levels of noise, thanks to our pre-processing, our network is able to blindly deal with arbitrary levels of noise and therefore is well-suited to deal with spatially variant noise levels which is quite common in modern MRIs. We trained the network using around 300.000 different patches at every epoch selected from the IXI dataset.

Once the network is trained, the test image is filtered using an overcomplete 3D sliding window approach. This overcomplete approach further reduces the noise by averaging several overlapping estimations and contributes to reduce block artifacts. Finally, the resulting noise image estimation is subtracted from the noisy input and the normalization is inverted (i.e. the image is multiplied by the local standard deviation image and the local mean image is added) to produce the final denoised image.

### *Rotational invariant denoising*

As shown in [7,8], when a *prefiltered* image is available we can use this image as a guide image for a rotationally invariant NLM filter to robustly perform a local similarity estimation defined as follow:

$$\hat{A}(i) = \frac{\sum_{j \in \Omega} w(i,j)\, y(i)}{\sum_{j \in \Omega} w(i,j)} \qquad w(i,j) = e^{-\frac{1}{2}\left( \frac{(g(i)-g(j))^2 + 3(\mu_{N_i} - \mu_{N_j})^2}{2h_i^2} \right)} \qquad (2)$$

where $\mu_{Ni}$ and $\mu_{Nj}$ are the mean values of patches *Ni* and *Nj* around voxels *i* and *j* in the guide image *g*, *h* is related to the standard deviation of the noise present on image *y* at voxel *i-th* and $\Omega$ represents position of the elements of the search volume. Full details of the rotational invariant NLM filter can be found in the original paper [7].

It is worth noting that applying this rotational invariant NLM using the proposed PBCNN guide image not only outperforms the use of the PBCNN filter only but also further minimizes the block artifact problem (note that the RI-NLM is applied to the original noisy image and not to the PBCNN denoised image). We call this combined filter as PRI-PBCNN for Prefiltered Rotational Invariant PBCNN.

### *Adaptation to Rician noise*

The asymmetry of the Rician distribution leads to a non-constant intensity bias depending on the local SNR. To reduce such bias, some authors have proposed to remove the bias in the squared magnitude image [37,3].

We have used a spatially varying version of this method to be able to deal with a wider range of situations. The Rician bias correction for PBCNN and PRI-PBCNN methods is performed in squared magnitude image as done in the original PRI-NLM method [7].

$$\hat{A}(i) = \sqrt{\max\left(\left(\frac{\sum_{j\in\Omega} w(i,j)\, y(i)^2}{\sum_{j\in\Omega} w(i,j)}\right) - 2\sigma(i)^2, 0\right)}$$

(3)

Where in this case the fixed bias $2\sigma(i)^2$ depends on the local noise level $\sigma(i)$ at position *i.* Local noise level is directly estimated from the noise map generated from the PBCNN method and corrected in the Rician case using the following lookup table approach.

### *Rician noise estimation*

Since MR images are typically corrupted with Rician noise, we need to estimate it to correct the bias it introduces in the denoised images. In Manjón et al [8] we presented a new mapping-based approach to estimate the original noise field – before the Rician correction – using the local standard deviation of the image residuals (difference between noisy and filtered images, the local standard deviation was estimated using a local region of 3x3x3 voxels).

Specifically, we estimated a mapping function correcting the systematic noise underestimation by using Monte-Carlo simulations where effective local SNR (that is the ratio between the local mean and the corresponding local standard deviation) was related to the correction factor that converts the local Gaussian-like standard deviation to the corresponding Rician one. We fitted the simulated data using the following rational model.

$$\Phi(\gamma) = \begin{cases} \dfrac{((0.9846(\gamma - 1.86) + 0.1983)}{((\gamma - 1.86) + 0.1175))} & if\ (\gamma > 1.86) \\ \qquad 0 \quad otherwise \end{cases}$$

(4)

where γ represents the effective local SNR. The corrected local standard deviation is calculated by multiplying the correction factor based on the effective local SNR with the initially estimated local standard deviation provided by the proposed PBCNN method.

$$\hat{\sigma} = \sigma \, \Phi(\gamma) \qquad (5)$$

Figure 2 shows a visual example of the estimated noise field with and without correction for stationary Rician noise. As can be noted, the used residual mapping method corrects the severe underestimation of the noise level at low intensity regions.

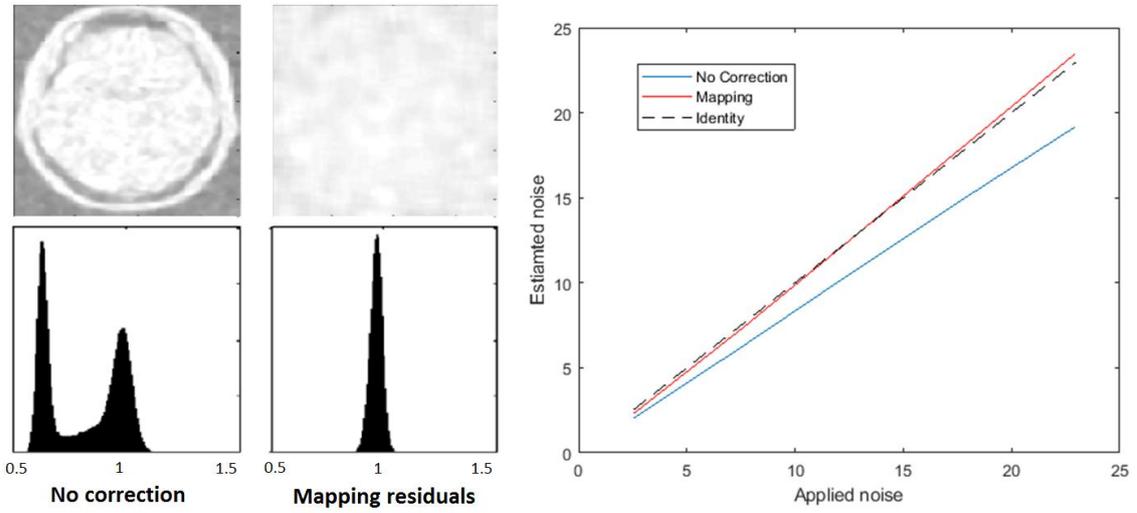

***Figure 2.*** *Brainweb example. Left: Example of stationary Rician noise estimation results (9%). The noise fields are shown in the [0,1] range after dividing the estimated noise field by the real noise field. From left to right. First column: PBCNN without Rician correction and corresponding histogram. Second column: PBCNN residual-based noise estimation with the proposed mapping correction and corresponding histogram. Right: Comparison of the two noise estimation methods for different noise levels (with and without correction). Uncorrected estimation severely underestimates noise levels as the noise level increases while corrected estimation provides an almost perfect estimation.*

Therefore, we used the estimated noise field to correct the Rician noise induce bias in the PBCNN denoised image and to initialize the following RI-NLM method. Finally, the full method consist in: image preprocessing, denoising using PBCNN, Rician bias correction and denoising using RI-NLM.

## 3. Results

In this section, a set of experiments show how the parameters of the proposed method were selected and present comparisons with state-of-the-art methods. Two quantitative measures were used to evaluate the results, the Peak Signal to Noise Ratio (PSNR) and the structural similarity index (SSIM) [38], which is a measure more consistent with the human visual system.

### 3.1. Network topology and parameter settings

We explored several options to design the proposed PBCNN, such as the patch-size, number of layers or the number of filters. For the number of filters, we tested 16, 32 and 64 filters, the results showed that the higher the number of filters the better the results. We chose 64 filters because 128 significantly increased the model size and training time but the improvement was modest. Regarding the number of layers, we found that increasing the number of layers to have a receptive field wider than the patch size was not improving significantly the results. We tested patch sizes of 6x6x6, 12x12x12 and 24x24x24 voxels and we found that the best results were obtained for 12x12x12 voxels (with 7 internal blocks covering a receptive field of 17x17x17 voxels). Besides, the use of small patches allows to denoise images with spatially varying noise fields more effectively.

We trained the designed network using the described IXI dataset. The final results were evaluated on the testing Brainweb dataset. We added Gaussian noise (range 1 to 9%) to the training images to simulate noisy cases. We did not add Rician noise as the Rician bias correction is performed at postprocessing as previously described.

### *Impact of Normalization and loss function*

We evaluate the use of different normalization layers and loss functions as they play a key role in the training process. For this experiment, we measured the validation metric (MSE) using 10000 training patches and 1000 validation patches randomly selected (MSE loss was used in this experiment). In figure 4 the comparison of Batch Normalization and Instance normalization is compared (batch size=128). As can be noted Instance normalization provides a faster convergence and a lower validation loss than BatchNormalization.

We also compared several loss functions using the same settings. Specifically, we compare the usual mean squared error (MSE), the mean absolute error (MAE) and a custom loss called Mix that is computed as the sum of MSE and MAE. We tried this Mix loss because we observed that MAE seems to work better than MSE for low levels of noise while MSE works better for high levels of noise. In figure 3 the effect of the different loss functions on the evolution of the validation loss. As can be noticed the proposed Mix loss has a slightly lower error than MSE and MAE and therefore was used as the loss function of the proposed method.

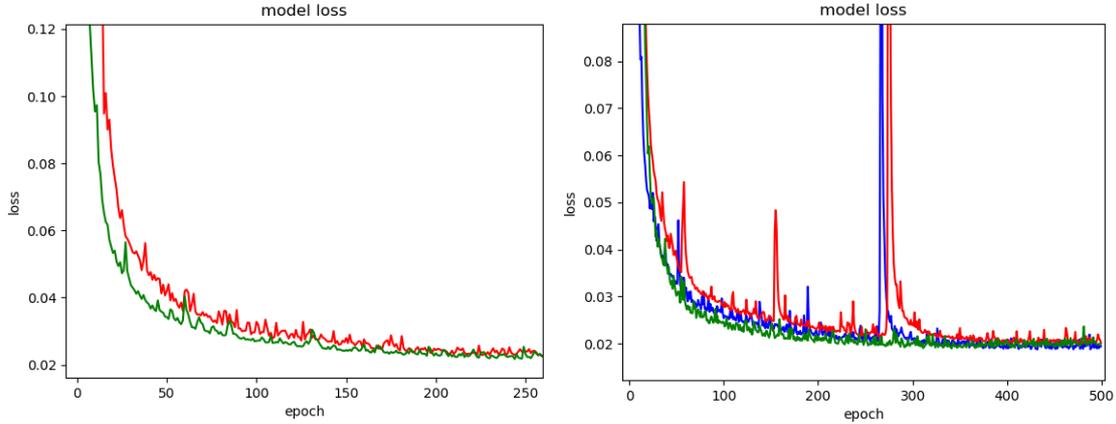

***Figure 3.*** *Left: Influence of the normalization layer. Red (BatchNormalization) and Green (InstanceNormalization). As can be noticed Instance Normalization provides a faster convergence and a lower loss than BatchNormalization. Right: Influence of the loss function. Red (MSE), Blue (Mix) and Green(MAE). Mix provided a slightly lower loss than MSE and MAE.*

### *Impact of overlapping*

We also evaluated the impact of the overlapping level over the final results (i.e., test dataset). To be consistent with previously published methods PSNR was measured only in the foreground area excluding the background (voxels higher than 10 in the Brainweb dataset). Specifically, offsets of 12, 6 and 3 voxels in all 3 dimensions were evaluated. The results are shown in table 1. We found also that the higher the overlap the better the results (at the expense of a higher computational time (2, 15 and 105 seconds)).

***Table 1****: PSNR results on the T1 Brainweb phantom for the proposed PBCNN method for stationary Gaussian noise with two different overlapping levels.*

| Filter | Noise level | | | | | |
|---|---|---|---|---|---|---|
| | **1%** | **3%** | **5%** | **7%** | **9%** | **Mean** |
| **Noisy** | 39.99 | 30.46 | 26.02 | 23.10 | 20.91 | 28.10 |
| **PBCNN (offset=12)** | 43.47 | 37.85 | 35.32 | 33.61 | 32.30 | 36.51 |
| **PBCNN (offset=6)** | 43.91 | 38.41 | 35.95 | 34.29 | 33.01 | 37.11 |
| **PBCNN (offset=3)** | **44.13** | **38.66** | **36.24** | **34.61** | **33.35** | **37.40** |

This fact is not surprising as this has been previously observed in previously proposed filters such as the NLPCA or ODCT filters among others [7,8]. We used an offset of 3 as default in our proposed method.

### PBCNN Noise estimation

We also evaluated the ability of the network to estimate the noise present in the images. After denoising we simply estimated the noise level calculating the standard deviation of the residuals generated subtracting the denoised image from the original noisy image. In table 2 the estimated noise for several levels of Gaussian and Rician noise is shown. As can be noted, the proposed method estimated the noise in both Gaussian and Rician case in a consistent manner.

**Table 2**: Comparison of the applied and estimated noise using the PBCNN method over Gaussian and Rician noise corrupted images.

| Noise Level | 1% | 3% | 5% | 7% | 9% |
|---|---|---|---|---|---|
| *Applied Noise* | 2.55 | 7.65 | 12.74 | 17.84 | 22.95 |
| *Estimated Noise (Gaussian)* | 2.35 | 7.24 | 12.25 | 17.28 | 22.35 |
| *Estimated Noise (Rician)* | 2.27 | 7.36 | 12.60 | 17.84 | 23.13 |

### 3.3. Methods comparison

We compared our proposed method with other recent state-of-the-art denoising methods. The methods used for comparisons were i) patch-based methods: NL-PCA and PRI-NL-PCA (Manjón el al., 2015), ABM4D [39], BM4D [40] and ii) deep learning methods: MCDnCNNg (blind) and MCDnCNNs (noise specific) [31]. Blind methods such as MCDnCNNg or the proposed PBCNN are trained using data of different noise levels while noise specific networks such MCDnCNNs are trained for a specific noise level and therefore it can be only applied optimally to images with this specific noise level which highly limits its applicability (and requires a previous noise estimation step). Both Gaussian and Rician noise with different levels were evaluated.

Table 3 shows the results on stationary Gaussian and Rician noise. As can be noticed, the combination of the proposed PBCNN with the RI-NLM provided similar results than the PRI-NLPCA method outperforming it for medium and high levels of noise. In figure 4 an example of the obtained results is shown.

**Table 3.** *PSNR and SSIM results (*computed over the brain (i.e. foreground))* for the compared methods applied on stationary noise (Gaussian and Rician).*

| Noise | Filter | Noise level | | | | | |
|---|---|---|---|---|---|---|---|
| | | 1% | 3% | 5% | 7% | 9% | Average |
| | **Noisy** | 39.99\|0.970 | 30.46\|0.814 | 26.02\|0.656 | 23.10\|0.530 | 20.91\|0.433 | 28.10\|0.681 |
| | **NLPCA** | 44.80\|**0.994** | 38.97\|0.979 | 36.40\|0.964 | 34.67\|0.948 | 33.32\|0.931 | 37.63\|0.963 |
| | **PRI-NLPCA** | **45.38**\|**0.994** | 39.33\|**0.981** | 36.63\|**0.968** | 34.90\|0.955 | 33.58\|0.941 | **37.96**\|**0.968** |
| **Gauss.** | **BM4D** | 44.02\|0.992 | 38.35\|0.975 | 35.91\|0.960 | 34.31\|0.945 | 33.10\|0.930 | 37.14\|0.960 |
| | **PBCNN** | 44.13\|0.993 | 38.66\|0.978 | 36.24\|0.964 | 34.61\|0.950 | 33.35\|0.936 | 37.40\|0.964 |
| | **PRI-PBCNN** | 45.09\|**0.994** | **39.34**\|**0.981** | **36.70**\|**0.968** | **35.00**\|**0.956** | **33.67**\|**0.942** | **37.96**\|**0.968** |
| | | | | | | | |
| | **Noisy** | 40.00\|0.970 | 30.49\|0.815 | 26.09\|0656 | 23.20\|0.529 | 21.04\|0.431 | 28.16\|0.680 |
| | **NLPCA** | 44.79\|**0.994** | 38.90\|0.978 | 36.23\|0.962 | 34.37\|0.943 | 32.88\|0.923 | 37.43\|0.960 |
| | **PRI-NLPCA** | **45.31**\|**0.994** | **39.34**\|**0.981** | 36.58\|**0.967** | 34.74\|0.952 | 33.28\|0.935 | **37.85**\|**0.966** |
| **Rician** | **BM4D** | 44.09\|0.992 | 38.35\|0.975 | 35.84\|0.959 | 34.17\|0.942 | 32.88\|0.924 | 36.99\|0.958 |
| | **PBCNN** | 44.12\|0.993 | 38.61\|0.977 | 36.13\|0.962 | 34.35\|0.946 | 32.91\|0.928 | 37.22\|0.961 |
| | **PRI-PBCNN** | 45.10\|**0.994** | 39.29\|**0.981** | **36.63**\|**0.967** | **34.84**\|**0.953** | **33.41**\|**0.937** | **37.85**\|**0.966** |

Table 4 shows the results on spatially varying Gaussian and Rician noise. In this case, we used a simulated noise field similar to those that can be found on parallel imaging. To generate the noise field a modulation map with factors 1 to 3 was multiplied with different levels of Rician noise (1% to 9%).

**Table 4.** *PSNR and SSIM results (*computed over the brain (i.e. foreground))* for the compared methods applied on spatially varying noise.*

| Noise | Filter | Noise Level | | | | | |
|---|---|---|---|---|---|---|---|
| | | 1-3% | 3-9% | 5-15% | 7-21% | 9-27% | Average |
| | **Noisy** | 34.34\|0.900 | 24.80\|0.621 | 20.36\|0.442 | 17.44\|0.328 | 15.26\|0.253 | 22.44\|0.508 |
| | **NLPCA** | 41.66\|0.987 | 35.95\|0.958 | 33.17\|0.925 | 31.31\|0.891 | 29.92\|0.857 | 34.40\|0.924 |
| | **PRI-NLPCA** | **42.25**\|**0.989** | 36.30\|0.965 | 33.52\|0.939 | 31.61\|0.911 | 30.16\|0.883 | 34.77\|0.937 |
| **Gauss.** | **ABM4D** | 40.45\|0.980 | 35.48\|0.960 | 33.10\|0.930 | 31.48\|0.900 | **30.24**\|0.870 | 34.15\|0.928 |
| | **PBCNN** | 41.30\|0.986 | 35.96\|0.961 | 33.39\|0.935 | 31.54\|0.907 | 30.10\|0.878 | 34.46\|0.933 |
| | **PRI-PBCNN** | 42.12\|**0.989** | **36.45**\|**0.966** | **33.73**\|**0.942** | **31.70**\|**0.915** | 30.04\|**0.885** | **34.81**\|**0.939** |
| | | | | | | | |
| | **Noisy** | 34.35\|0.900 | 24.87\|0.621 | 20.50\|0.441 | 17.64\|0.325 | 15.50\|0.247 | 22.57\|0.507 |
| | **NL-PCA** | 41.64\|0.987 | 35.77\|0.956 | 32.74\|0.917 | 30.48\|0.873 | 28.42\|0.824 | 33.81\|0.911 |
| | **PRI-NL-PCA** | **42.23**\|**0.989** | 36.19\|0.964 | 33.15\|0.934 | **30.87**\|**0.897** | **28.83**\|**0.856** | **34.25**\|**0.928** |
| **Rician** | **ABM4D** | 40.43\|0.980 | 34.41\|0.940 | 31.27\|0.890 | 28.80\|0.820 | 26.55\|0.740 | 32.29\|0.874 |
| | **PBCNN** | 41.28\|0.986 | 35.83\|0.959 | 32.86\|0.918 | 30.27\|0.871 | 27.71\|0.819 | 33.59\|0.916 |
| | **PRI-PBCNN** | 42.11\|**0.989** | **36.31**\|**0.965** | **33.20**\|**0.935** | 30.49\|0.896 | 27.84\|0.821 | 33.99\|0.922 |

In this case, the combination of the proposed PBCNN with the RI-NLM provided the best overall results for Gaussian noise but not for Rician noise were the PRI-NLPCA provided slightly better results.

Note though that the processing time of the PRI-NLPCA is around 8 minutes while it was under the 2 minutes using the proposed PRI-PBCNN (this time can be further reduced below 1 minute by reducing the patch overlap in the PCNN part with a minimal drop in the filtering quality).

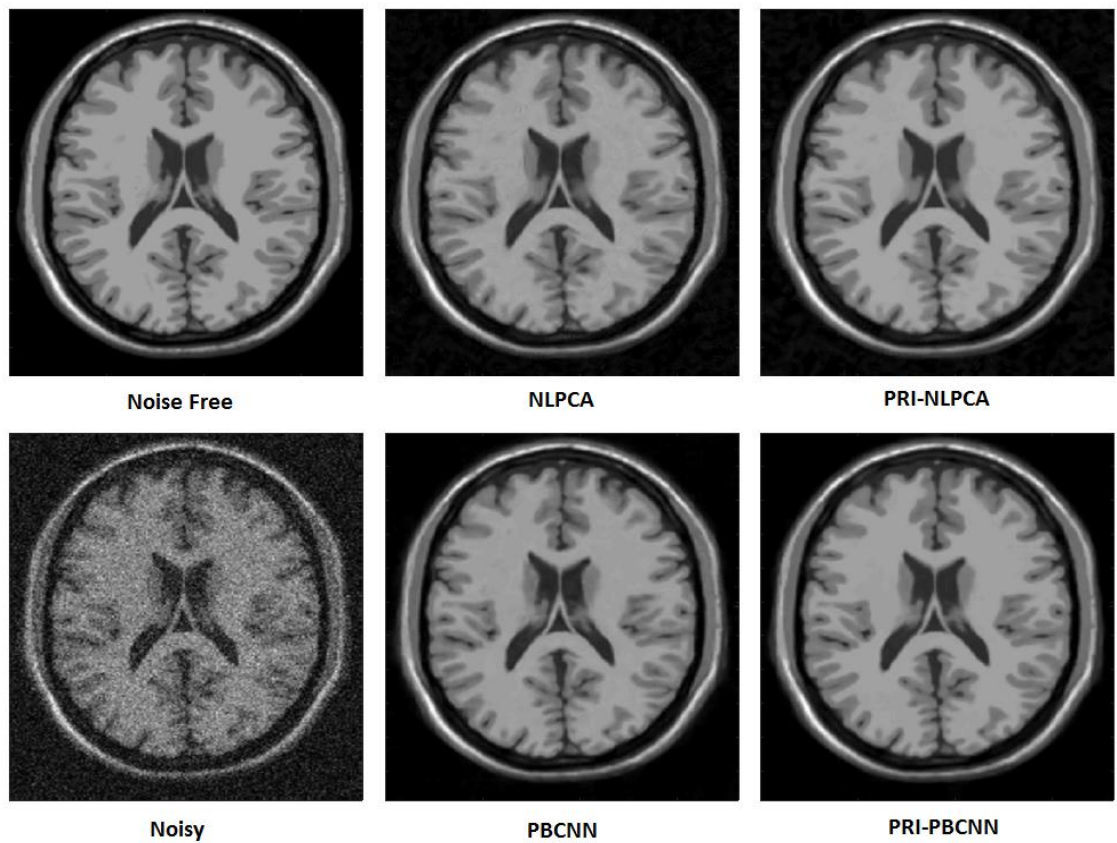

*Figure 4. Example results of the compared filters for 9% of Rician noise. Note that the proposed PBCNN result is very similar to the NLPCA method but NLPCA shows some truncation artifacts. As can noticed the proposed PRI-PBCNN method provides the best results.*

We also compared our results with other previously proposed CNN-based methods. Specifically, we compare our proposed method with the MCDnCNNg (blind) and MCDnCNNs (noise specific) [31] methods. Note that in this case the PSNR was computed over the whole volume (not only in the foreground) and only for stationary Rician noise since the authors measured the PSNR in this way. Results are presented in table 5.

**Table 5.** *PSNR results of the Deep learning compared methods for stationary Rician noise (*computed over the whole volume (i.e. background and foreground)). *Results of* MCDnCNNg *and* MCDnCNNg *methods were estimated from figure 6 in the paper.*

| Noise | Filter | Noise level | | | | | |
|---|---|---|---|---|---|---|---|
| | | 1% | 3% | 5% | 7% | 9% | Average |
| | **Noisy** | 38.89 | 29.19 | 24.69 | 21.73 | 19.53 | 26.81 |
| | **MCDnCNNg** | 43.80 | 38.10| | 36.00| | 34.50 | 33.00 | 37.08 |
| **Rician** | **MCDnCNNs** | **45.20** | **39.80** | 37.20 | 35.20 | 33.50 | 38.18 |
| | **PBCNN** | 44.82 | 39.50 | 37.29 | 35.71 | 34.53 | 38.37 |
| | **PRI-PBCNN** | 44.98 | 39.48 | **37.64** | **36.54** | **35.33** | **38.80** |

The proposed PBCNN method outperformed the blind MCDnCNNg method by a large margin (more than 1 dB) for all noise levels and also provided a better overall performance than the MCDnCNNs method. The proposed PRI-PBCNN further improved the results.

## 3.4. Qualitative evaluation on real images

Although the results on synthetic data are easy to interpret, in some cases they might be not realistic enough as they are simplifications of the real images. To qualitatively evaluate the results of the proposed method we applied it to two real images (where no noise was added but that were acquired at a resolution close to 0.5 mm isotropic) and we evaluated visually the results. In figure 5, the results can be visually checked. As can be noticed, no anatomical information can be observed in the residuals (a part from the Rician induced bias) and the residual map shows the spatially varying nature of the noise present in both images.

Finally, it is worth noting that a network trained on T1 images can be used to denoise T2 images effectively because the networks learns how to remove edges from anatomical image no matter the contrast. This is a really interesting feature as the proposed filter can be blindly used to denoise any kind of anatomical MR images (not only T1).

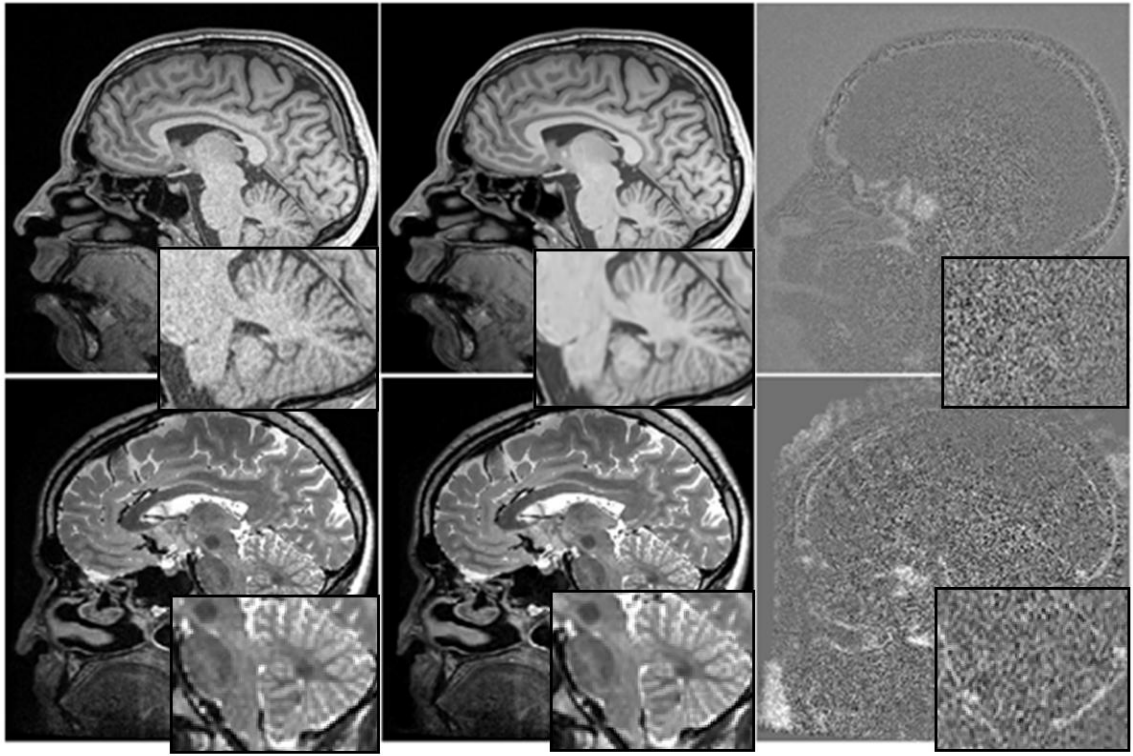

**Figure 5**: *Denoising example of real T1 and T2 images. From left to right: Noisy image, filtered image with the proposed filter and residual image (removed Rician noise). A close-up of the cerebellum of the T1 image is shown to better see the denoising effect.*

## 4. Conclusion

In this paper, we presented a new method for MRI denoising that combines the benefits of new deep learning techniques with the strength of the traditional non-local image processing methods. The proposed method is based on an overcomplete patch-based CNN which produces a *prefiltered* image that is used as a guide image within a rotational invariant non-local means framework.

The proposed patch-based CNN benefits from use of instance normalization layers instead of classical batch normalization layers by using individual patch properties instead of using the minibatch variance to normalize the individual patch features. On the other hand, the use of an overcomplete approach significantly improved the quality of the filtering and can be seen as an ensemble approach to leverage individual results.

The proposed method performed similar than PRI-NLPCA method for low and medium noise levels and outperformed it for high noise levels (for both Gaussian and Rician noise) and is an effective approach to automatically reduce the amount on noise in MR images in a blind manner thanks to its automatic adaptation to different levels of noise. Furthermore, method is able to deal with spatially variant noise (as can be noticed on table 4 and figure 4) thanks to the local variance standardization and its adaptive patch-based nature. From an efficiency point of view the proposed method is 4 times faster than the PRI-NLPCA method with around 2 minutes processing time. However, this time can be significantly reduced using a lower patch overlap, going from minutes to seconds with a small drop in the denoising quality.

When compared to other Deep learning methods proposed for MRI denoising, we found that our method outperformed similar blind version while performed better than the non-blind version for medium and high levels of noise. The lower results for low noise levels may come from the fact that the networks pay more attention to samples with high levels of noise as the loss gain is higher on those. We will explore in the future an adaptive network bank trained on different SNR conditions to better extract the noise from the images in some specific conditions.

Finally, differently from other deep learning approaches, it is worth to note that the proposed method can be applied to any kind of MR images (T1, T2, FLAIR, etc) because it has been designed to estimate noise present in the images independently of the local patterns present on them which makes it ideal to be used in a wide range of research and clinical settings.

## Acknowledgements

We want to thank Thomas Tourdias and Canon Medical for providing the high-resolution datasets. This research was supported by the Spanish DPI2017-87743-R grant from the Ministerio de Economia, Industria y Competitividad of Spain. This study has been also carried out with financial support from the French State, managed by the French National Research Agency (ANR) in the frame of the Investments for the future Pro-gram IdEx Bordeaux (ANR-10-IDEX-03-02, HL-MRI Project) and Cluster of excellence CPU and TRAIL (HR-DTI ANR-10-LABX-57). The authors gratefully acknowledge the support of NVIDIA Corporation with their donation of the TITAN X GPU used in this research.